\documentclass[10pt,aps,prl,twocolumn,preprintnumbers,nofootinbib]{revtex4-2}
\usepackage{amsmath,amssymb,mathtools}
\usepackage{hyperref}
\usepackage{xspace}
\usepackage{graphicx}
\usepackage{grffile}
\usepackage{enumitem}
\newcommand{\sect}[1]{%
\bigskip\noindent%
{\bfseries\upshape\rmfamily\boldmath{#1}.}---%
\ignorespaces%
}

\newcommand{\Qavg}{Q_{\rm avg}}
\newcommand{\der}{\mathrm{d}}

\newcommand{\Fcal}{\mathcal{F}}
 \newcommand{\Ccal}{\mathcal{C}}
\newcommand{\hx}{\hat{x}}
\newcommand{\hz}{\hat{z}}
\newcommand{\qb}{\bar{q}}
\newcommand{\qp}{q^{\prime}}
\newcommand{\qbp}{\bar{q}^{\prime}}
\newcommand{\GeV}{\mathrm{GeV}}
\newcommand{\code}[1]{\texttt{#1}}

\begin{document}

\preprint{ZU-TH 07/24}

\title{Semi-inclusive deep-inelastic scattering at NNLO in QCD}%

\author{Leonardo Bonino}
\affiliation{%
  Physik-Institut, Universit\"at Z\"urich, Winterthurerstrasse 190, 8057 Z\"urich, Switzerland}%
\author{Thomas Gehrmann}
 \affiliation{%
  Physik-Institut, Universit\"at Z\"urich, Winterthurerstrasse 190, 8057 Z\"urich, Switzerland}%
\author{Giovanni Stagnitto}
\affiliation{%
Universit\`{a} degli Studi di Milano-Bicocca \& INFN, Piazza della Scienza 3, 20216 Milano, Italy}%


\begin{abstract}
  Semi-inclusive hadron production processes in deep-inelastic lepton-nucleon
  scattering are important probes of the quark flavour structure of the nucleon
  and of the fragmentation dynamics of quarks into hadrons.  We compute the full
  next-to-next-to-leading order (NNLO) QCD corrections to the coefficient
  functions for semi-inclusive deep-inelastic scattering (SIDIS) in analytical
  form.  The numerical impact of these corrections for precision physics is
  illustrated by a detailed comparison with data on single inclusive hadron
  spectra from the CERN COMPASS experiment.
\end{abstract}

\maketitle

\section{Introduction}

Identified hadron production in hard scattering processes is described in
quantum chromodynamics (QCD) through the production of partons (quarks or
gluons) which subsequently fragment into hadrons. This parton-to-hadron
transition is a non-perturbative process that can be parametrized in terms of
fragmentation functions (FFs) which describe the probability of a parton
fragmenting into a hadron carrying some fraction of its
momentum~\cite{Field:1976ve,Field:1977fa}. These FFs fulfil Altarelli-Parisi
evolution equations in their resolution scale~\cite{Altarelli:1977zs}, which are
in complete analogy to the evolution of parton distributions functions (PDFs) in
the nucleon.

Production cross sections for a variety of hadron species have been measured in
electron-positron, lepton-hadron and hadron-hadron collisions. To include these
data sets into a global fit of FFs requires the knowledge of the respective
parton-level coefficient functions (differential in the momentum of the
fragmenting parton) to the desired perturbative order. At present, these
coefficient functions are known to next-to-leading order (NLO) for
lepton-hadron~\cite{Altarelli:1979kv,Baier:1979sp,deFlorian:1997zj} and
hadron-hadron collisions~\cite{Aversa:1988vb} and to next-to-next-to-leading
order (NNLO) for $e^+e^-$ annihilation~\cite{Rijken:1996ns,Mitov:2006ic}.
Consequently, global fits of FFs at
NNLO~\cite{Bertone:2017tyb,Anderle:2015lqa,Abdolmaleki:2021yjf} focus only on
data from $e^+e^-$ experiments, while having to discard any other collider data.

Semi-inclusive hadron production in deep-inelastic lepton-nucleon scattering
(SIDIS) has been measured
extensively~\cite{EuropeanMuon:1991sne,ZEUS:1995acw,H1:1996muf,HERMES:2012uyd,COMPASS:2016xvm}
for various hadrons.  By considering different hadron species, it is possible to
single out different flavour combinations of incoming partons, thereby probing
the detailed quark and antiquark flavour decomposition of the PDFs. This SIDIS
information is largely complementary to inclusive DIS structure function
measurements, which allow to determine only a single flavour combination to high
accuracy. Moreover, SIDIS measurements play an important role in the
determination of spin-dependent PDFs~\cite{deFlorian:2009vb} that have to rely
on far fewer hadron-collider observables than ordinary (spin-averaged) PDFs.
Again, these studies can be performed in a self-consistent manner only up to NLO
due to the unavailability of corrections to the SIDIS coefficient functions at
higher orders.

It is the purpose of this work to enable precision physics studies with SIDIS
observables by deriving the full NNLO QCD corrections to the SIDIS coefficient
functions. We provide their analytical expressions for all partonic channels
(combinations of initial state and identified final-state partons) and study the
impact of the newly derived corrections on a representative data set on SIDIS
charged pion production from COMPASS~\cite{COMPASS:2016xvm}.

\section{Kinematics of SIDIS}

We consider the observation of a hadron $h$ following the scattering of a lepton
on a nucleon. We closely follow the notation of~\cite{Anderle:2012rq},
describing semi-inclusive deep-inelastic scattering as
$\ell(k)\,p(P)\to\ell(k^{\prime})\,h(P_h)\,X$, with $X$ inclusive final-state
radiation. The leptons $\ell$ momenta define the four-momentum $q=k-k^{\prime}$
of the exchanged virtual vector boson and the energy transfer $y=(P\cdot
q)/(P\cdot k)$.  The usual exclusive variables for $Q^2=-q^2$,
\begin{align}
x=\frac{Q^2}{2P\cdot q}\, ,\quad \quad z=\frac{P\cdot P_h}{P\cdot q}  \,  ,
\end{align}
describe the momentum fraction of the nucleon carried by the incoming parton
($x$) and the momentum fraction of the outgoing parton carried by the identified
hadron ($z$).  For $\sqrt{s}$ center-of-mass energy of the lepton-nucleon system
we have $Q^2=xys$.

As we consider values of $Q\ll M_Z$ only (highly) virtual photons are exchanged,
and the triple-differential cross section reads
\begin{eqnarray}\label{d3sigdxdydz}
  \frac{\der^3\sigma^h}{\der x \der y \der z} &=& \frac{4\pi\alpha^2}{Q^2} \bigg[
    \frac{1+(1-y)^2}{2y} \Fcal^h_T(x,z,Q^2)
    \nonumber \\ && \phantom{\frac{4\pi\alpha^2}{Q^2} \bigg[}
    + \frac{1-y}{y} \Fcal^h_L(x,z,Q^2) \bigg] \, ,
\end{eqnarray}
with $\alpha$ denoting the fine structure constant. 

The transverse $\Fcal^h_T$ and longitudinal $\Fcal^h_L$ SIDIS structure
functions are given by the sum over all partonic channels of the convolution
between the PDF for a parton $p$ ($f_p$), the FF of a parton $p^{\prime}$ into
the hadron $h$ ($D^h_{p^{\prime}}$), and the coefficient function for the
transition $p\to p^{\prime}$ ($\Ccal^i_{p' p}$):
\begin{align}
  \Fcal_i^h(x,z,Q^2) &= \sum_{p,p'} \int_x^1 \frac{\der\hx}{\hx}
  \int_z^1 \frac{\der\hz}{\hz} f_p\left(\frac{x}{\hx},\mu_F^2\right)
  D_{p'}^h\left(\frac{z}{\hz},\mu_A^2\right)  
  \nonumber \\ 
  &\times\,\Ccal^i_{p' p}\left(\hx,\hz,Q^2,\mu_R^2,\mu_F^2,\mu_A^2\right)\,,
  \quad i = T,L \, .
\end{align}
The factorisation theorem that allows the above expression introduces two
factorisation scales: $\mu_F$ for the initial state and $\mu_A$ for the final
state.  With $\mu_R$ we indicate the renormalisation scale.  The coefficient
functions encode the hard-scattering part of the process and can be computed in
perturbative QCD. Their perturbative expansion in the strong coupling constant
$\alpha_s$ reads
\begin{equation}
  \Ccal^i_{p' p} = C^{i,(0)}_{p' p}
  + \frac{\alpha_s(\mu_R^2)}{2\pi} C^{i,(1)}_{p' p}
  + \left(\frac{\alpha_s(\mu_R^2)}{2\pi}\right)^2  C^{i,(2)}_{p' p}
  + \mathcal{O}(\alpha_s^3)\, .
\end{equation}

At LO, only the $qq$ channel ($\gamma^* q\to q$) contributes, with the LO
coefficient functions trivially given by
\begin{equation}
  C^{T,(0)}_{qq} = e_q^2 \delta(1-\hx) \delta(1-\hz)\,,\quad
  C^{L,(0)}_{qq} = 0\,,
\end{equation}
where $e_q$ is the quark's charge.  At NLO instead also the channels $qg$ and
$gq$ start to contribute, and results for $C^{i,(1)}_{qq}$, $C^{i,(1)}_{gq}$ and
$C^{i,(1)}_{qg}$ can be found in the literature~\cite{Anderle:2012rq}.

In this Letter we present results for the NNLO corrections
$C^{i,(2)}_{p^{\prime}p}$ to all partonic channels appearing at this order.
Following the notation of~\cite{Anderle:2016kwa}, the $7$ partonic channels
appearing at $\mathcal{O}(\alpha_s^2)$ are:
\begin{align}\label{CFNNLOlist}
C^{i,(2)}_{qq}&=C^{i,(2)}_{\qb\qb}=e_q^2 C^{i,\mathrm{NS}}_{qq}+\biggl( \sum_j e^2_{q_j}\biggr)C^{i,\mathrm{PS}}_{qq} \, , \nonumber \\
C^{i,(2)}_{\qb q}&=C^{i,(2)}_{q \qb}=e_q^2C^{i}_{\qb q} \, , \nonumber \\
C^{i,(2)}_{\qp q}&=C^{i,(2)}_{\qbp \qb}=e_q^2 C^{i,1}_{\qp q}+e_{\qp}^2 C^{i,2}_{\qp q}+e_q e_{\qp} C^{i,3}_{\qp q} \, , \nonumber \\
C^{i,(2)}_{\qbp q}&=C^{i,(2)}_{\qp \qb}=e_q^2 C^{i,1}_{\qp q}+e_{\qp}^2 C^{i,2}_{\qp q}-e_q e_{\qp} C^{i,3}_{\qp q} \, , \nonumber \\
C^{i,(2)}_{gq}&=C^{i,(2)}_{g \qb}=e_q^2 C^{i}_{gq} \, , \nonumber \\
C^{i,(2)}_{qg}&=C^{i,(2)}_{\qb g}=e_q^2 C^{i}_{qg} \, , \nonumber \\
C^{i,(2)}_{gg}&=\biggl( \sum_j e_{q_j}^2 \biggr) C^{i}_{gg} \,  ,
\end{align}
again for $i=T,L$. With $\qp$ ($\qbp$) we indicate a quark (antiquark) of
flavour different from $q$, whereas the NS and PS superscripts in the quark-to-quark channel denote the non-singlet and the pure-singlet components respectively.

The coefficient functions are computed by applying projectors to extract the
longitudinal and transverse components from the respective parton-level
subprocess matrix elements with incoming kinematics fixed by $Q^2$ and
$\hat{x}$, which are then integrated over the final state phase space. This
integration is fully inclusive in the extra radiation $X$ and keeps the final
state momentum fraction of the parton $p'$ fixed to $\hat{z}$.

\section{Method}

\begin{figure}[t]
  \includegraphics[width=0.2\textwidth]{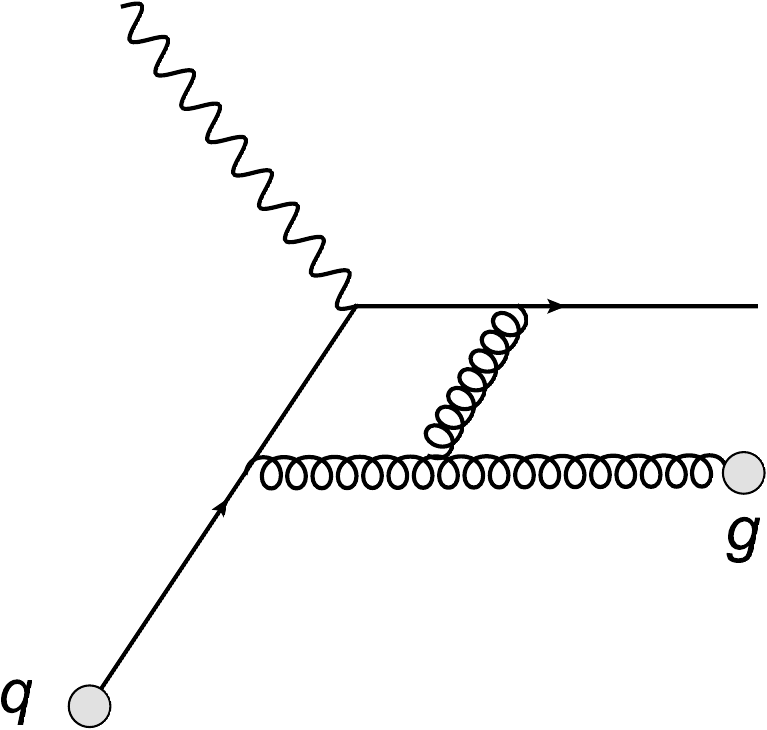}\\
  \includegraphics[width=0.2\textwidth]{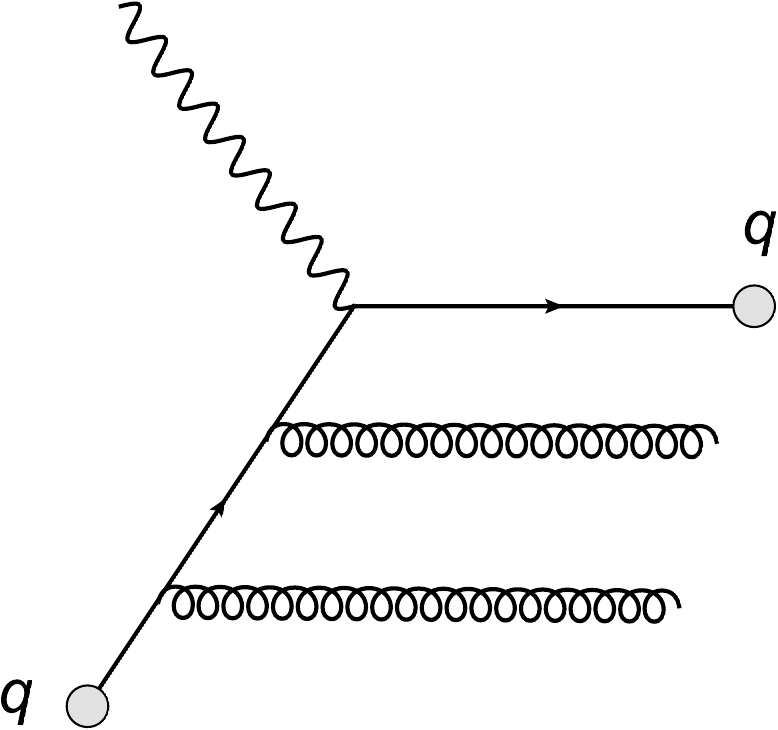}\,
  \includegraphics[width=0.2\textwidth]{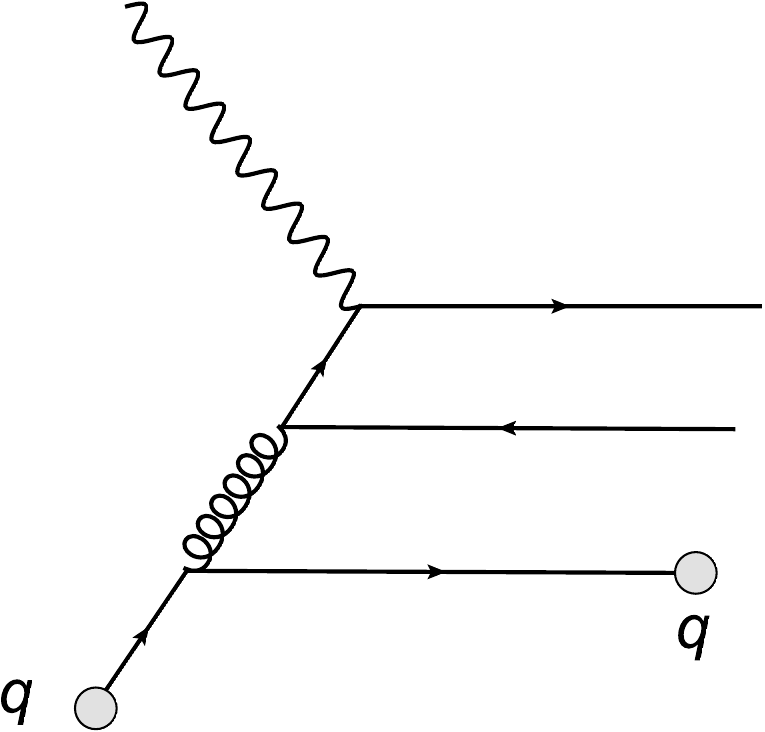}
  \caption{Example Feynman diagrams contributing to $C^i_{gq}$ at the RV level
    (top), and to $C^{i,\mathrm{NS}}_{qq}$ and $C^{i,\mathrm{PS}}_{qq}$ at the RR
    level (bottom left and right).}
  \label{fig:fdiag}
\end{figure}

At NNLO in QCD, three types of parton-level contributions must be taken into
account, relative to the underlying Born-level process: two-loop virtual
corrections (double-virtual, VV), one-loop corrections to single real radiation
processes (real-virtual, RV) and tree-level double real radiation processes
(RR), with example diagrams shown in Figure~\ref{fig:fdiag}. These are
accompanied by contributions from QCD renormalization and mass factorization of
the PDFs and FFs.

NNLO QCD corrections to processes with identified particles at hadron colliders
have recently been derived for identified
photons~\cite{Gehrmann:2022cih,Chen:2022gpk} and for the production of bottom
hadrons in top quark decays~\cite{Czakon:2022pyz}. These calculations are
performed in a fully exclusive manner in the form of parton-level event
generators which provide the full kinematical information on all final state
partons (and on a single identified hadron or photon), which can then be
subjected to the precise final state definition that is used in the
experiment. To enable these computations, a method to identify and extract
infrared singular real radiation up to NNLO had to be employed. The calculation
for identified photons used the antenna subtraction
method~\cite{Gehrmann-DeRidder:2005btv,Daleo:2006xa,Currie:2013vh}, while the
bottom hadron production relies on a sector-improved residue
subtraction~\cite{Czakon:2010td}.
   
The analytic ingredients to the antenna subtraction method are so-called antenna
functions, which encapsulate all infrared singular radiation that emerges
between two hard radiator partons. These antenna functions are used to construct
the real radiation subtraction terms, and they are integrated analytically over
the respective antenna phase spaces to make the infrared pole structure
explicit.  In the case of fragmentation processes~\cite{Gehrmann:2022pzd}, one
of the hard radiators is the fragmenting parton, while the other can be in the
initial or in the final state. The kinematical situation of fragmentation
antenna functions with one radiator in the initial state corresponds exactly to
the kinematics of SIDIS~\cite{Gehrmann:2021lwb}. 
Consequently, the RV and RR contributions to SIDIS can
be obtained employing exactly the same methods as were used to derive the
integrated fragmentation antenna functions with one initial state
radiator. These methods were described in detail in~\cite{Gehrmann:2022cih} and
we only provide a brief summary here.

The one-loop squared matrix elements for the RV contributions can be expressed
in terms of one-loop bubble and box integrals, which are known in exact form in
$\epsilon$,
with $\epsilon$ the dimensional regularisation parameter
for $N = 4-2\epsilon$ space-time dimensions.
For fixed $\hat{x}$ and $\hat{z}$, the associated phase space
integral is fully constrained, such that only expansions in the end-point
distributions in $\hat{x}=1$ and $\hat{z}=1$ are required to obtain the final
result for this contribution. To avoid ambiguities associated with the analytic
continuation of the one-loop master integrals, the parameter space of the RV
contribution is segmented into four sectors: $(\hat{x}\leq 0.5, \hat{x}\leq
\hat{z}\leq 1-\hat{x})$, $(\hat{z}\leq 0.5, \hat{z}<\hat{x}\leq 1-\hat{z})$,
$(\hat{x}>0.5,1-\hat{x}< \hat{z}<\hat{x})$ and
$(\hat{z}>0.5,1-\hat{z}<\hat{x}\leq \hat{z})$, where manifestly real-valued
expressions for the contributions are obtained~\cite{Gehrmann:2022cih}. The
expressions are continuous across the boundaries of the regions.

The RR contributions correspond to integrations over a three-particle phase
space, with the momentum fraction of one of the particles fixed by
$\hat{z}$. They can be expressed as cuts of two-loop integrals in forward
kinematics, with $\hat{z}$ expressed as a linear cut propagator. These integrals
are reduced to master integrals using integration-by-parts (IBP)
identities~\cite{Chetyrkin:1981qh,Laporta:2000dsw}, as implemented in
\code{Reduze2}~\cite{vonManteuffel:2012np}.  The RR contributions to the SIDIS
coefficient functions are expressed in terms of 13 integral families, which
contain a total of 21 master integrals. These master integrals are determined by
solving their differential equations~\cite{Gehrmann:1999as} in $\hat{x}$ and
$\hat{z}$, using \code{PolyLogTools}~\cite{Duhr:2019tlz} and
\code{HPL}~\cite{Maitre:2005uu}. The boundary terms for these differential
equations are obtained by integrating the generic solutions over $\hat{z}$ and
comparing to the master integrals relevant to inclusive integrated antenna
functions~\cite{Daleo:2009yj} with one initial-state and one final-state
radiator.  Of the 21 master integrals, 9 were already computed in the context of
photon fragmentation at NNLO~\cite{Gehrmann:2022cih}, and derivation of the
remaining 12 integrals will be described in detail elsewhere~\cite{IFfrag}.

The VV contributions correspond to the well-known two-loop quark form
factor~\cite{Gehrmann:2005pd} in space-like kinematics.

All contributions are computed using \code{FORM}~\cite{Vermaseren:2000nd} and
are assembled to yield the bare SIDIS coefficient functions, which still contain
ultraviolet and collinear pole terms. By adding the renormalization and mass
factorization counterterms (including convolutions of lower-order terms using
\code{MT}~\cite{Hoschele:2013pvt}), the finite physical SIDIS coefficient
functions are obtained.

\section{Results}

\begin{figure*}[t]
  \includegraphics[width=0.9\textwidth]{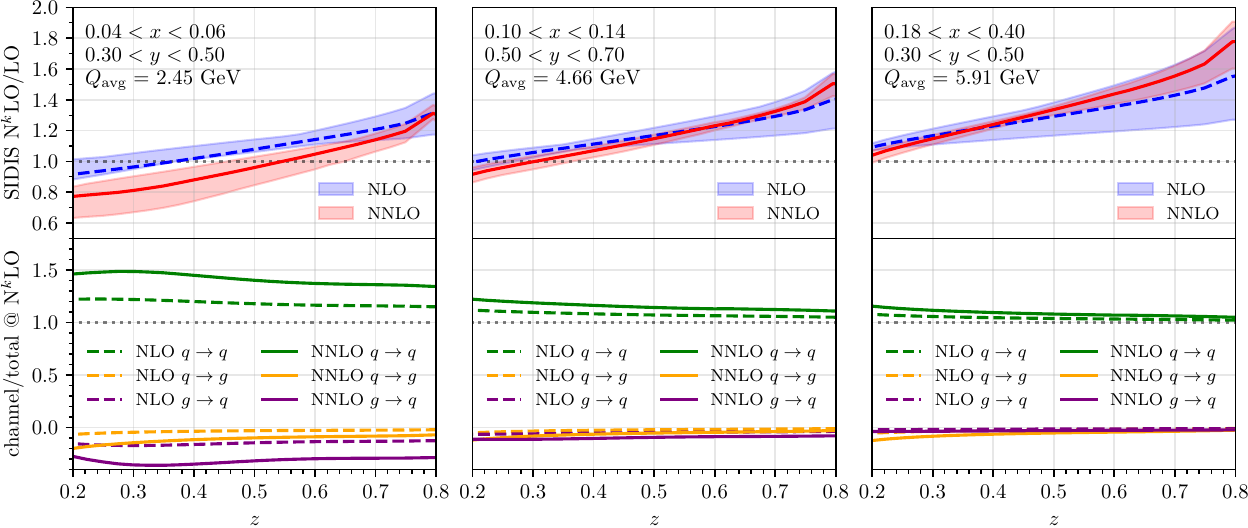}
  \caption{QCD $K$-factors up to NNLO and fractional contribution of individual
    channels for selected kinematical bins studied by the COMPASS
    experiment~\protect\cite{COMPASS:2016xvm}. The $g \to g$, $q \to \bar{q}$,
    $q \to \qp$ and $q \to \qbp$ channels are not shown in the channel
    decomposition as they are found to give negligible contributions in the
    kinematical bins considered.}
  \label{fig:kfact}
\end{figure*}

The results for the full set of coefficient functions up to NNLO order are too
lengthy to be presented here and thus are given as an ancillary file attached to
the \code{arXiv} submission of this Letter.  Our results include the full scale
dependence ($\mu_R$, $\mu_F$, and $\mu_A$) that was cross-checked with the
solution of the renormalisation group equation for all channels.  In this
section we discuss the comparison of our results with the literature and
numerical results.

By means of the threshold resummation formalism for SIDIS~\cite{Anderle:2012rq},
approximate corrections for the $qq$ channel have been derived at
NNLO~\cite{Abele:2021nyo}, and even at N$^3$LO~\cite{Abele:2022wuy}.
Such approximate NNLO corrections have been adopted in the context of a global
QCD analysis of light fragmentation
functions~\cite{Borsa:2022vvp,AbdulKhalek:2022laj}.
Moreover, partial results for the $qq$ NNLO longitudinal coefficient function
are also available~\cite{Anderle:2016kwa}.
Most recently the leading colour contribution to the $qq$ non-singlet channel
was computed in \cite{Goyal:2023xfi}.

Concerning the $qq$ channel, we compare our results against the ones
of~\cite{Goyal:2023xfi}.  The longitudinal components are in perfect numerical
agreement.  Regarding the leading color transverse ones, we find analytical
agreement for all terms involving endpoint distributions as well as perfect
numerical agreement for the regular part in the region $(\hat{z}\leq 0.5,
\hat{z}<\hat{x}\leq 1-\hat{z})$.  We are also in agreement with the threshold
expansion terms of~\cite{Abele:2021nyo}, which predict all double distributions
in the partonic variables, and have been confirmed by~\cite{Goyal:2023xfi} as
well.

Figure~\ref{fig:kfact} illustrates the numerical impact of the newly computed
NNLO corrections and assesses the relevance of different partonic channels.
Using selected kinematical bins from the COMPASS
measurement~\cite{COMPASS:2016xvm} of SIDIS pion production in muon-nucleon
scattering, which is described in more detail in the following section, we
compute the $K$-factors at NLO and NNLO and decompose the cross sections
according to different channels. We use the NNPDF3.1 PDF
set~\cite{NNPDF:2017mvq} and the FF set from~\cite{Borsa:2022vvp} at NNLO
throughout, with $\alpha_s(M_Z)=0.118$ and with $N_F = 5$ light quarks. The
central scales are fixed at $\mu_R=\mu_F=\mu_A=Q$, with scale variations
determined through variations by a factor 2 around the central scale.  We
further fix $\mu_F=\mu_A$.

We observe moderate NNLO corrections to the $K$-factors, which reinforce the
tendency of the NLO corrections of an increase of the $K$-factor with increasing
$z$. The non-uniformity of the NNLO corrections in $x$ and $z$ clearly
highlights the phenomenological relevance of the NNLO contributions.  In the
smallest $x$-bin (corresponding to the lowest $\Qavg$), NNLO corrections are
somewhat larger, and the overlap of NLO and NNLO uncertainty bands is only
marginal. At larger $x$, the NNLO corrections are within the NLO uncertainty
bands and their inclusion leads to considerably smaller uncertainties from 20\%
at NLO to well below 10\% at NNLO.  The predictions are largely dominated by the
quark-to-quark channel, the gluon-to-quark and quark-to-gluon channels both
yield small negative corrections to the SIDIS cross section, especially at small
$x$, with the gluon-to-quark channel being typically larger due to the larger
magnitude of the respective fragmentation function.  All new channels appearing
at NNLO are found to give negligible contributions.

\section{Comparison with data}\label{sec:CWD}

The COMPASS experiment performs deep-inelastic scattering measurements on
various fixed targets with a high-energy muon beam at CERN.  In their SIDIS
study~\cite{COMPASS:2016xvm}, momentum spectra for charged pions and for
unidentified charged hadrons are measured with a 160~GeV muon beam scattering off
an isoscalar target, corresponding to a center of mass energy $\sqrt{s} \approx
17.35\, \GeV$.  Events are accepted if $Q^2>1\,\GeV^2$ and $W>5\,\GeV$ with
$W=\sqrt{(P+q)^2}$ the invariant mass of the hadronic system.

\begin{figure*}[t]
\includegraphics[width=\textwidth]{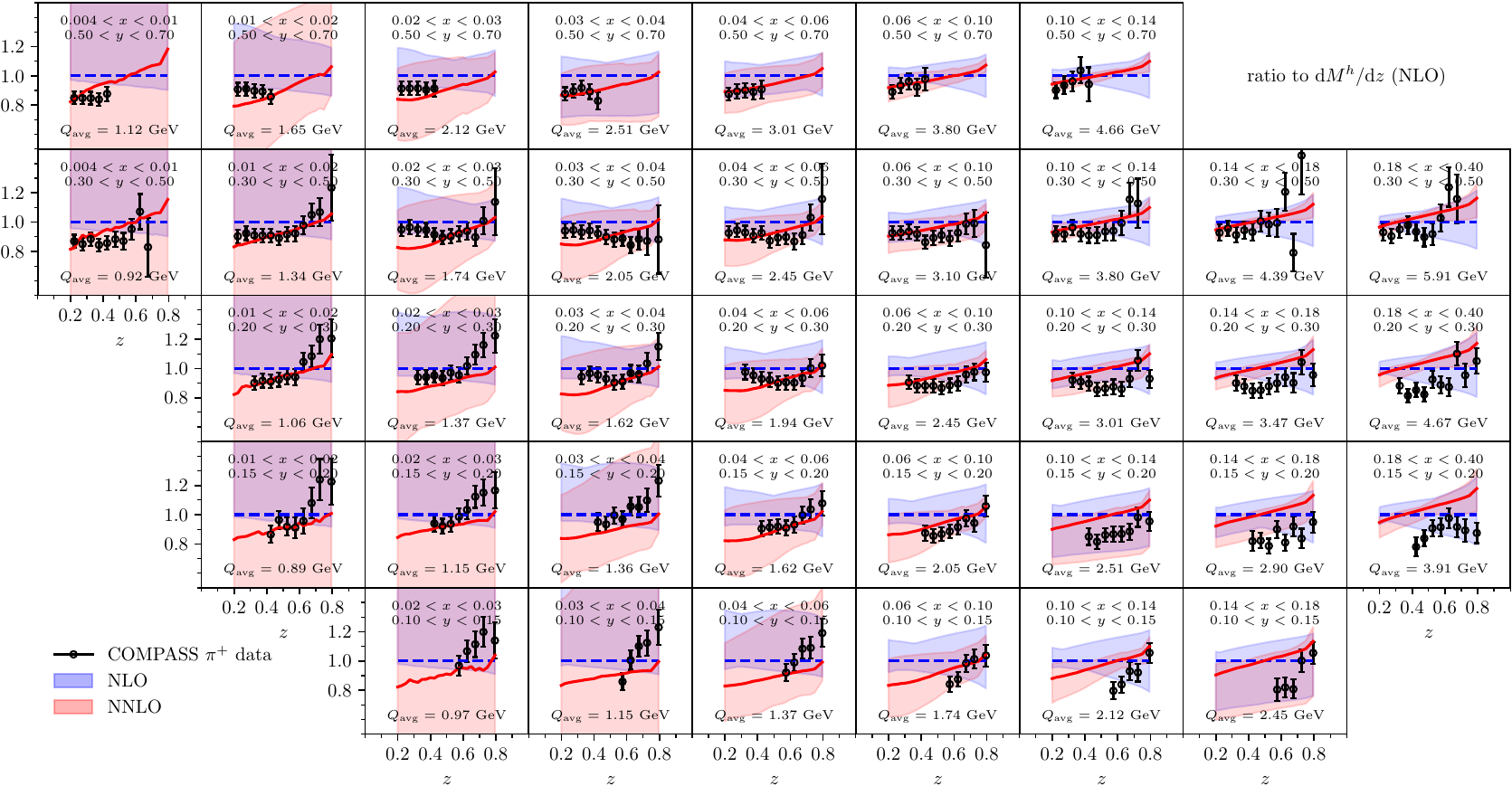}
\caption{Comparison of NLO and NNLO theory predictions for $\pi^+$ multiplicities with COMPASS data.}
\label{fig:compass}
\end{figure*}

The measured hadron multiplicities $\der M^h/\der z$ are given by the ratio of
the differential cross section for hadron production and the differential
inclusive DIS cross section. Therefore we compute the ratio
\begin{align}\label{eq:RhSIDIS}
\frac{\der M^h}{\der z} = \frac{\der^3 \sigma^h/\der x\der y \der z}{\der^2 \sigma/\der x\der y} \, .
\end{align} 
For brevity, we focus on the $h=\pi^+$ spectra.  In our numerical implementation
we compute the denominator of~\eqref{eq:RhSIDIS} using the \code{APFEL++}
code~\cite{Bertone:2013vaa,Bertone:2017gds}.  We apply the same experimental
cuts in our numerical implementation and we integrate~\eqref{eq:RhSIDIS} over
$x$ and $y$, according to the given bin ranges. In Figure~\ref{fig:compass} we
present the ratio of data and theory predictions over the NLO result.  The
uncertainty on theory predictions is estimated by varying the scales in an
uncorrelated way between the numerator and denominator of~\eqref{eq:RhSIDIS}.

We observe that inclusion of the NNLO corrections modifies the shape of the
predictions, in general improving the description of the experimental data.  For
the lowest values of $\Qavg \leq 2$~GeV, no reduction of the scale uncertainty
is observed. Moving to higher $\Qavg$, this reduction becomes clearly
significant, with NNLO uncertainties usually being half the size of their NLO
counterparts.

\section{Conclusions}
Semi-inclusive deep inelastic scattering processes will be among the key
observables of the physics program at the BNL Electron-Ion Collider (EIC).
To enable precision studies
with SIDIS data, higher order perturbative corrections are crucial. To prepare
the precision SIDIS program at EIC, we derived the analytical expressions for
the NNLO QCD corrections to the SIDIS coefficient functions. The NNLO
corrections are non-uniform in the kinematical variables. They lead to a
substantial reduction of the uncertainty on the theory predictions at
sufficiently large values of $Q$, where the perturbative expansion is
applicable. In comparison with COMPASS results on $\pi^+$ SIDIS production, we
observe an improved description of the experimental data.

Our newly derived results allow precision determinations of the quark flavour
decomposition of nucleon PDFs and of hadron FFs in SIDIS at the EIC. A natural
extension of our work could be towards the polarized SIDIS coefficient
functions, thereby enabling precision SIDIS studies in the EIC spin physics
program.

\sect{Acknowledgements}
We are grateful to Kay Sch\"onwald, Markus L\"ochner and Valerio Bertone for
discussions and for checks on the results, to Robin Sch\"urmann for early
contributions to the calculations described here and to Daniel de Florian for
discussions and for providing us with the FF set of~\cite{Borsa:2022vvp}.
This work has received funding from the Swiss National Science Foundation (SNF)
under contract 200020-204200 and from the European Research Council (ERC) under
the European Union's Horizon 2020 research and innovation programme grant
agreement 101019620 (ERC Advanced Grant TOPUP).

\bibliography{sidis}

\begin{thebibliography}{48}%
\makeatletter
\providecommand \@ifxundefined [1]{%
 \@ifx{#1\undefined}
}%
\providecommand \@ifnum [1]{%
 \ifnum #1\expandafter \@firstoftwo
 \else \expandafter \@secondoftwo
 \fi
}%
\providecommand \@ifx [1]{%
 \ifx #1\expandafter \@firstoftwo
 \else \expandafter \@secondoftwo
 \fi
}%
\providecommand \natexlab [1]{#1}%
\providecommand \enquote  [1]{``#1''}%
\providecommand \bibnamefont  [1]{#1}%
\providecommand \bibfnamefont [1]{#1}%
\providecommand \citenamefont [1]{#1}%
\providecommand \href@noop [0]{\@secondoftwo}%
\providecommand \href [0]{\begingroup \@sanitize@url \@href}%
\providecommand \@href[1]{\@@startlink{#1}\@@href}%
\providecommand \@@href[1]{\endgroup#1\@@endlink}%
\providecommand \@sanitize@url [0]{\catcode `\\12\catcode `\$12\catcode
  `\&12\catcode `\#12\catcode `\^12\catcode `\_12\catcode `\%12\relax}%
\providecommand \@@startlink[1]{}%
\providecommand \@@endlink[0]{}%
\providecommand \url  [0]{\begingroup\@sanitize@url \@url }%
\providecommand \@url [1]{\endgroup\@href {#1}{\urlprefix }}%
\providecommand \urlprefix  [0]{URL }%
\providecommand \Eprint [0]{\href }%
\providecommand \doibase [0]{http://dx.doi.org/}%
\providecommand \selectlanguage [0]{\@gobble}%
\providecommand \bibinfo  [0]{\@secondoftwo}%
\providecommand \bibfield  [0]{\@secondoftwo}%
\providecommand \translation [1]{[#1]}%
\providecommand \BibitemOpen [0]{}%
\providecommand \bibitemStop [0]{}%
\providecommand \bibitemNoStop [0]{.\EOS\space}%
\providecommand \EOS [0]{\spacefactor3000\relax}%
\providecommand \BibitemShut  [1]{\csname bibitem#1\endcsname}%
\let\auto@bib@innerbib\@empty
\bibitem [{\citenamefont {Field}\ and\ \citenamefont
  {Feynman}(1977)}]{Field:1976ve}%
  \BibitemOpen
  \bibfield  {author} {\bibinfo {author} {\bibfnamefont {R.~D.}\ \bibnamefont
  {Field}}\ and\ \bibinfo {author} {\bibfnamefont {R.~P.}\ \bibnamefont
  {Feynman}},\ }\href {\doibase 10.1103/PhysRevD.15.2590} {\bibfield  {journal}
  {\bibinfo  {journal} {Phys. Rev. D}\ }\textbf {\bibinfo {volume} {15}},\
  \bibinfo {pages} {2590} (\bibinfo {year} {1977})}\BibitemShut {NoStop}%
\bibitem [{\citenamefont {Field}\ and\ \citenamefont
  {Feynman}(1978)}]{Field:1977fa}%
  \BibitemOpen
  \bibfield  {author} {\bibinfo {author} {\bibfnamefont {R.~D.}\ \bibnamefont
  {Field}}\ and\ \bibinfo {author} {\bibfnamefont {R.~P.}\ \bibnamefont
  {Feynman}},\ }\href {\doibase 10.1016/0550-3213(78)90015-9} {\bibfield
  {journal} {\bibinfo  {journal} {Nucl. Phys. B}\ }\textbf {\bibinfo {volume}
  {136}},\ \bibinfo {pages} {1} (\bibinfo {year} {1978})}\BibitemShut {NoStop}%
\bibitem [{\citenamefont {Altarelli}\ and\ \citenamefont
  {Parisi}(1977)}]{Altarelli:1977zs}%
  \BibitemOpen
  \bibfield  {author} {\bibinfo {author} {\bibfnamefont {G.}~\bibnamefont
  {Altarelli}}\ and\ \bibinfo {author} {\bibfnamefont {G.}~\bibnamefont
  {Parisi}},\ }\href {\doibase 10.1016/0550-3213(77)90384-4} {\bibfield
  {journal} {\bibinfo  {journal} {Nucl. Phys. B}\ }\textbf {\bibinfo {volume}
  {126}},\ \bibinfo {pages} {298} (\bibinfo {year} {1977})}\BibitemShut
  {NoStop}%
\bibitem [{\citenamefont {Altarelli}\ \emph {et~al.}(1979)\citenamefont
  {Altarelli}, \citenamefont {Ellis}, \citenamefont {Martinelli},\ and\
  \citenamefont {Pi}}]{Altarelli:1979kv}%
  \BibitemOpen
  \bibfield  {author} {\bibinfo {author} {\bibfnamefont {G.}~\bibnamefont
  {Altarelli}}, \bibinfo {author} {\bibfnamefont {R.~K.}\ \bibnamefont
  {Ellis}}, \bibinfo {author} {\bibfnamefont {G.}~\bibnamefont {Martinelli}}, \
  and\ \bibinfo {author} {\bibfnamefont {S.-Y.}\ \bibnamefont {Pi}},\ }\href
  {\doibase 10.1016/0550-3213(79)90062-2} {\bibfield  {journal} {\bibinfo
  {journal} {Nucl. Phys. B}\ }\textbf {\bibinfo {volume} {160}},\ \bibinfo
  {pages} {301} (\bibinfo {year} {1979})}\BibitemShut {NoStop}%
\bibitem [{\citenamefont {Baier}\ and\ \citenamefont
  {Fey}(1979)}]{Baier:1979sp}%
  \BibitemOpen
  \bibfield  {author} {\bibinfo {author} {\bibfnamefont {R.}~\bibnamefont
  {Baier}}\ and\ \bibinfo {author} {\bibfnamefont {K.}~\bibnamefont {Fey}},\
  }\href {\doibase 10.1007/BF01545897} {\bibfield  {journal} {\bibinfo
  {journal} {Z. Phys. C}\ }\textbf {\bibinfo {volume} {2}},\ \bibinfo {pages}
  {339} (\bibinfo {year} {1979})}\BibitemShut {NoStop}%
\bibitem [{\citenamefont {de~Florian}\ \emph {et~al.}(1998)\citenamefont
  {de~Florian}, \citenamefont {Stratmann},\ and\ \citenamefont
  {Vogelsang}}]{deFlorian:1997zj}%
  \BibitemOpen
  \bibfield  {author} {\bibinfo {author} {\bibfnamefont {D.}~\bibnamefont
  {de~Florian}}, \bibinfo {author} {\bibfnamefont {M.}~\bibnamefont
  {Stratmann}}, \ and\ \bibinfo {author} {\bibfnamefont {W.}~\bibnamefont
  {Vogelsang}},\ }\href {\doibase 10.1103/PhysRevD.57.5811} {\bibfield
  {journal} {\bibinfo  {journal} {Phys. Rev. D}\ }\textbf {\bibinfo {volume}
  {57}},\ \bibinfo {pages} {5811} (\bibinfo {year} {1998})},\ \Eprint
  {http://arxiv.org/abs/hep-ph/9711387} {arXiv:hep-ph/9711387} \BibitemShut
  {NoStop}%
\bibitem [{\citenamefont {Aversa}\ \emph {et~al.}(1989)\citenamefont {Aversa},
  \citenamefont {Chiappetta}, \citenamefont {Greco},\ and\ \citenamefont
  {Guillet}}]{Aversa:1988vb}%
  \BibitemOpen
  \bibfield  {author} {\bibinfo {author} {\bibfnamefont {F.}~\bibnamefont
  {Aversa}}, \bibinfo {author} {\bibfnamefont {P.}~\bibnamefont {Chiappetta}},
  \bibinfo {author} {\bibfnamefont {M.}~\bibnamefont {Greco}}, \ and\ \bibinfo
  {author} {\bibfnamefont {J.~P.}\ \bibnamefont {Guillet}},\ }\href {\doibase
  10.1016/0550-3213(89)90288-5} {\bibfield  {journal} {\bibinfo  {journal}
  {Nucl. Phys. B}\ }\textbf {\bibinfo {volume} {327}},\ \bibinfo {pages} {105}
  (\bibinfo {year} {1989})}\BibitemShut {NoStop}%
\bibitem [{\citenamefont {Rijken}\ and\ \citenamefont {van
  Neerven}(1997)}]{Rijken:1996ns}%
  \BibitemOpen
  \bibfield  {author} {\bibinfo {author} {\bibfnamefont {P.~J.}\ \bibnamefont
  {Rijken}}\ and\ \bibinfo {author} {\bibfnamefont {W.~L.}\ \bibnamefont {van
  Neerven}},\ }\href {\doibase 10.1016/S0550-3213(96)00669-4} {\bibfield
  {journal} {\bibinfo  {journal} {Nucl. Phys. B}\ }\textbf {\bibinfo {volume}
  {487}},\ \bibinfo {pages} {233} (\bibinfo {year} {1997})},\ \Eprint
  {http://arxiv.org/abs/hep-ph/9609377} {arXiv:hep-ph/9609377} \BibitemShut
  {NoStop}%
\bibitem [{\citenamefont {Mitov}\ \emph {et~al.}(2006)\citenamefont {Mitov},
  \citenamefont {Moch},\ and\ \citenamefont {Vogt}}]{Mitov:2006ic}%
  \BibitemOpen
  \bibfield  {author} {\bibinfo {author} {\bibfnamefont {A.}~\bibnamefont
  {Mitov}}, \bibinfo {author} {\bibfnamefont {S.}~\bibnamefont {Moch}}, \ and\
  \bibinfo {author} {\bibfnamefont {A.}~\bibnamefont {Vogt}},\ }\href {\doibase
  10.1016/j.physletb.2006.05.005} {\bibfield  {journal} {\bibinfo  {journal}
  {Phys. Lett. B}\ }\textbf {\bibinfo {volume} {638}},\ \bibinfo {pages} {61}
  (\bibinfo {year} {2006})},\ \Eprint {http://arxiv.org/abs/hep-ph/0604053}
  {arXiv:hep-ph/0604053} \BibitemShut {NoStop}%
\bibitem [{\citenamefont {Bertone}\ \emph {et~al.}(2017)\citenamefont
  {Bertone}, \citenamefont {Carrazza}, \citenamefont {Hartland}, \citenamefont
  {Nocera},\ and\ \citenamefont {Rojo}}]{Bertone:2017tyb}%
  \BibitemOpen
  \bibfield  {author} {\bibinfo {author} {\bibfnamefont {V.}~\bibnamefont
  {Bertone}}, \bibinfo {author} {\bibfnamefont {S.}~\bibnamefont {Carrazza}},
  \bibinfo {author} {\bibfnamefont {N.~P.}\ \bibnamefont {Hartland}}, \bibinfo
  {author} {\bibfnamefont {E.~R.}\ \bibnamefont {Nocera}}, \ and\ \bibinfo
  {author} {\bibfnamefont {J.}~\bibnamefont {Rojo}} (\bibinfo {collaboration}
  {NNPDF}),\ }\href {\doibase 10.1140/epjc/s10052-017-5088-y} {\bibfield
  {journal} {\bibinfo  {journal} {Eur. Phys. J. C}\ }\textbf {\bibinfo {volume}
  {77}},\ \bibinfo {pages} {516} (\bibinfo {year} {2017})},\ \Eprint
  {http://arxiv.org/abs/1706.07049} {arXiv:1706.07049 [hep-ph]} \BibitemShut
  {NoStop}%
\bibitem [{\citenamefont {Anderle}\ \emph {et~al.}(2015)\citenamefont
  {Anderle}, \citenamefont {Ringer},\ and\ \citenamefont
  {Stratmann}}]{Anderle:2015lqa}%
  \BibitemOpen
  \bibfield  {author} {\bibinfo {author} {\bibfnamefont {D.~P.}\ \bibnamefont
  {Anderle}}, \bibinfo {author} {\bibfnamefont {F.}~\bibnamefont {Ringer}}, \
  and\ \bibinfo {author} {\bibfnamefont {M.}~\bibnamefont {Stratmann}},\ }\href
  {\doibase 10.1103/PhysRevD.92.114017} {\bibfield  {journal} {\bibinfo
  {journal} {Phys. Rev. D}\ }\textbf {\bibinfo {volume} {92}},\ \bibinfo
  {pages} {114017} (\bibinfo {year} {2015})},\ \Eprint
  {http://arxiv.org/abs/1510.05845} {arXiv:1510.05845 [hep-ph]} \BibitemShut
  {NoStop}%
\bibitem [{\citenamefont {Abdolmaleki}\ \emph {et~al.}(2021)\citenamefont
  {Abdolmaleki}, \citenamefont {Soleymaninia}, \citenamefont {Khanpour},
  \citenamefont {Amoroso}, \citenamefont {Giuli}, \citenamefont {Glazov},
  \citenamefont {Luszczak}, \citenamefont {Olness},\ and\ \citenamefont
  {Zenaiev}}]{Abdolmaleki:2021yjf}%
  \BibitemOpen
  \bibfield  {author} {\bibinfo {author} {\bibfnamefont {H.}~\bibnamefont
  {Abdolmaleki}}, \bibinfo {author} {\bibfnamefont {M.}~\bibnamefont
  {Soleymaninia}}, \bibinfo {author} {\bibfnamefont {H.}~\bibnamefont
  {Khanpour}}, \bibinfo {author} {\bibfnamefont {S.}~\bibnamefont {Amoroso}},
  \bibinfo {author} {\bibfnamefont {F.}~\bibnamefont {Giuli}}, \bibinfo
  {author} {\bibfnamefont {A.}~\bibnamefont {Glazov}}, \bibinfo {author}
  {\bibfnamefont {A.}~\bibnamefont {Luszczak}}, \bibinfo {author}
  {\bibfnamefont {F.}~\bibnamefont {Olness}}, \ and\ \bibinfo {author}
  {\bibfnamefont {O.}~\bibnamefont {Zenaiev}},\ }\href {\doibase
  10.1103/PhysRevD.104.056019} {\bibfield  {journal} {\bibinfo  {journal}
  {Phys. Rev. D}\ }\textbf {\bibinfo {volume} {104}},\ \bibinfo {pages}
  {056019} (\bibinfo {year} {2021})},\ \Eprint
  {http://arxiv.org/abs/2105.11306} {arXiv:2105.11306 [hep-ph]} \BibitemShut
  {NoStop}%
\bibitem [{\citenamefont {Ashman}\ \emph {et~al.}(1991)\citenamefont {Ashman}
  \emph {et~al.}}]{EuropeanMuon:1991sne}%
  \BibitemOpen
  \bibfield  {author} {\bibinfo {author} {\bibfnamefont {J.}~\bibnamefont
  {Ashman}} \emph {et~al.} (\bibinfo {collaboration} {EMC}),\ }\href {\doibase
  10.1007/BF01559431} {\bibfield  {journal} {\bibinfo  {journal} {Z. Phys. C}\
  }\textbf {\bibinfo {volume} {52}},\ \bibinfo {pages} {361} (\bibinfo {year}
  {1991})}\BibitemShut {NoStop}%
\bibitem [{\citenamefont {Derrick}\ \emph {et~al.}(1996)\citenamefont {Derrick}
  \emph {et~al.}}]{ZEUS:1995acw}%
  \BibitemOpen
  \bibfield  {author} {\bibinfo {author} {\bibfnamefont {M.}~\bibnamefont
  {Derrick}} \emph {et~al.} (\bibinfo {collaboration} {ZEUS}),\ }\href
  {\doibase 10.1007/s002880050075} {\bibfield  {journal} {\bibinfo  {journal}
  {Z. Phys. C}\ }\textbf {\bibinfo {volume} {70}},\ \bibinfo {pages} {1}
  (\bibinfo {year} {1996})},\ \Eprint {http://arxiv.org/abs/hep-ex/9511010}
  {arXiv:hep-ex/9511010} \BibitemShut {NoStop}%
\bibitem [{\citenamefont {Adloff}\ \emph {et~al.}(1997)\citenamefont {Adloff}
  \emph {et~al.}}]{H1:1996muf}%
  \BibitemOpen
  \bibfield  {author} {\bibinfo {author} {\bibfnamefont {C.}~\bibnamefont
  {Adloff}} \emph {et~al.} (\bibinfo {collaboration} {H1}),\ }\href {\doibase
  10.1016/S0550-3213(96)00675-X} {\bibfield  {journal} {\bibinfo  {journal}
  {Nucl. Phys. B}\ }\textbf {\bibinfo {volume} {485}},\ \bibinfo {pages} {3}
  (\bibinfo {year} {1997})},\ \Eprint {http://arxiv.org/abs/hep-ex/9610006}
  {arXiv:hep-ex/9610006} \BibitemShut {NoStop}%
\bibitem [{\citenamefont {Airapetian}\ \emph {et~al.}(2013)\citenamefont
  {Airapetian} \emph {et~al.}}]{HERMES:2012uyd}%
  \BibitemOpen
  \bibfield  {author} {\bibinfo {author} {\bibfnamefont {A.}~\bibnamefont
  {Airapetian}} \emph {et~al.} (\bibinfo {collaboration} {HERMES}),\ }\href
  {\doibase 10.1103/PhysRevD.87.074029} {\bibfield  {journal} {\bibinfo
  {journal} {Phys. Rev. D}\ }\textbf {\bibinfo {volume} {87}},\ \bibinfo
  {pages} {074029} (\bibinfo {year} {2013})},\ \Eprint
  {http://arxiv.org/abs/1212.5407} {arXiv:1212.5407 [hep-ex]} \BibitemShut
  {NoStop}%
\bibitem [{\citenamefont {Adolph}\ \emph {et~al.}(2017)\citenamefont {Adolph}
  \emph {et~al.}}]{COMPASS:2016xvm}%
  \BibitemOpen
  \bibfield  {author} {\bibinfo {author} {\bibfnamefont {C.}~\bibnamefont
  {Adolph}} \emph {et~al.} (\bibinfo {collaboration} {COMPASS}),\ }\href
  {\doibase 10.1016/j.physletb.2016.09.042} {\bibfield  {journal} {\bibinfo
  {journal} {Phys. Lett. B}\ }\textbf {\bibinfo {volume} {764}},\ \bibinfo
  {pages} {1} (\bibinfo {year} {2017})},\ \Eprint
  {http://arxiv.org/abs/1604.02695} {arXiv:1604.02695 [hep-ex]} \BibitemShut
  {NoStop}%
\bibitem [{\citenamefont {de~Florian}\ \emph {et~al.}(2009)\citenamefont
  {de~Florian}, \citenamefont {Sassot}, \citenamefont {Stratmann},\ and\
  \citenamefont {Vogelsang}}]{deFlorian:2009vb}%
  \BibitemOpen
  \bibfield  {author} {\bibinfo {author} {\bibfnamefont {D.}~\bibnamefont
  {de~Florian}}, \bibinfo {author} {\bibfnamefont {R.}~\bibnamefont {Sassot}},
  \bibinfo {author} {\bibfnamefont {M.}~\bibnamefont {Stratmann}}, \ and\
  \bibinfo {author} {\bibfnamefont {W.}~\bibnamefont {Vogelsang}},\ }\href
  {\doibase 10.1103/PhysRevD.80.034030} {\bibfield  {journal} {\bibinfo
  {journal} {Phys. Rev. D}\ }\textbf {\bibinfo {volume} {80}},\ \bibinfo
  {pages} {034030} (\bibinfo {year} {2009})},\ \Eprint
  {http://arxiv.org/abs/0904.3821} {arXiv:0904.3821 [hep-ph]} \BibitemShut
  {NoStop}%
\bibitem [{\citenamefont {Anderle}\ \emph {et~al.}(2013)\citenamefont
  {Anderle}, \citenamefont {Ringer},\ and\ \citenamefont
  {Vogelsang}}]{Anderle:2012rq}%
  \BibitemOpen
  \bibfield  {author} {\bibinfo {author} {\bibfnamefont {D.~P.}\ \bibnamefont
  {Anderle}}, \bibinfo {author} {\bibfnamefont {F.}~\bibnamefont {Ringer}}, \
  and\ \bibinfo {author} {\bibfnamefont {W.}~\bibnamefont {Vogelsang}},\ }\href
  {\doibase 10.1103/PhysRevD.87.034014} {\bibfield  {journal} {\bibinfo
  {journal} {Phys. Rev. D}\ }\textbf {\bibinfo {volume} {87}},\ \bibinfo
  {pages} {034014} (\bibinfo {year} {2013})},\ \Eprint
  {http://arxiv.org/abs/1212.2099} {arXiv:1212.2099 [hep-ph]} \BibitemShut
  {NoStop}%
\bibitem [{\citenamefont {Anderle}\ \emph {et~al.}(2017)\citenamefont
  {Anderle}, \citenamefont {de~Florian},\ and\ \citenamefont
  {Rotstein~Habarnau}}]{Anderle:2016kwa}%
  \BibitemOpen
  \bibfield  {author} {\bibinfo {author} {\bibfnamefont {D.}~\bibnamefont
  {Anderle}}, \bibinfo {author} {\bibfnamefont {D.}~\bibnamefont {de~Florian}},
  \ and\ \bibinfo {author} {\bibfnamefont {Y.}~\bibnamefont
  {Rotstein~Habarnau}},\ }\href {\doibase 10.1103/PhysRevD.95.034027}
  {\bibfield  {journal} {\bibinfo  {journal} {Phys. Rev. D}\ }\textbf {\bibinfo
  {volume} {95}},\ \bibinfo {pages} {034027} (\bibinfo {year} {2017})},\
  \Eprint {http://arxiv.org/abs/1612.01293} {arXiv:1612.01293 [hep-ph]}
  \BibitemShut {NoStop}%
\bibitem [{\citenamefont {Gehrmann}\ and\ \citenamefont
  {Sch\"urmann}(2022{\natexlab{a}})}]{Gehrmann:2022cih}%
  \BibitemOpen
  \bibfield  {author} {\bibinfo {author} {\bibfnamefont {T.}~\bibnamefont
  {Gehrmann}}\ and\ \bibinfo {author} {\bibfnamefont {R.}~\bibnamefont
  {Sch\"urmann}},\ }\href {\doibase 10.1007/JHEP04(2022)031} {\bibfield
  {journal} {\bibinfo  {journal} {JHEP}\ }\textbf {\bibinfo {volume} {04}},\
  \bibinfo {pages} {031} (\bibinfo {year} {2022}{\natexlab{a}})},\ \Eprint
  {http://arxiv.org/abs/2201.06982} {arXiv:2201.06982 [hep-ph]} \BibitemShut
  {NoStop}%
\bibitem [{\citenamefont {Chen}\ \emph {et~al.}(2022)\citenamefont {Chen},
  \citenamefont {Gehrmann}, \citenamefont {Glover}, \citenamefont {H\"ofer},
  \citenamefont {Huss},\ and\ \citenamefont {Sch\"urmann}}]{Chen:2022gpk}%
  \BibitemOpen
  \bibfield  {author} {\bibinfo {author} {\bibfnamefont {X.}~\bibnamefont
  {Chen}}, \bibinfo {author} {\bibfnamefont {T.}~\bibnamefont {Gehrmann}},
  \bibinfo {author} {\bibfnamefont {E.~W.~N.}\ \bibnamefont {Glover}}, \bibinfo
  {author} {\bibfnamefont {M.}~\bibnamefont {H\"ofer}}, \bibinfo {author}
  {\bibfnamefont {A.}~\bibnamefont {Huss}}, \ and\ \bibinfo {author}
  {\bibfnamefont {R.}~\bibnamefont {Sch\"urmann}},\ }\href {\doibase
  10.1007/JHEP08(2022)094} {\bibfield  {journal} {\bibinfo  {journal} {JHEP}\
  }\textbf {\bibinfo {volume} {08}},\ \bibinfo {pages} {094} (\bibinfo {year}
  {2022})},\ \Eprint {http://arxiv.org/abs/2205.01516} {arXiv:2205.01516
  [hep-ph]} \BibitemShut {NoStop}%
\bibitem [{\citenamefont {Czakon}\ \emph {et~al.}(2023)\citenamefont {Czakon},
  \citenamefont {Generet}, \citenamefont {Mitov},\ and\ \citenamefont
  {Poncelet}}]{Czakon:2022pyz}%
  \BibitemOpen
  \bibfield  {author} {\bibinfo {author} {\bibfnamefont {M.}~\bibnamefont
  {Czakon}}, \bibinfo {author} {\bibfnamefont {T.}~\bibnamefont {Generet}},
  \bibinfo {author} {\bibfnamefont {A.}~\bibnamefont {Mitov}}, \ and\ \bibinfo
  {author} {\bibfnamefont {R.}~\bibnamefont {Poncelet}},\ }\href {\doibase
  10.1007/JHEP03(2023)251} {\bibfield  {journal} {\bibinfo  {journal} {JHEP}\
  }\textbf {\bibinfo {volume} {03}},\ \bibinfo {pages} {251} (\bibinfo {year}
  {2023})},\ \Eprint {http://arxiv.org/abs/2210.06078} {arXiv:2210.06078
  [hep-ph]} \BibitemShut {NoStop}%
\bibitem [{\citenamefont {Gehrmann-De~Ridder}\ \emph
  {et~al.}(2005)\citenamefont {Gehrmann-De~Ridder}, \citenamefont {Gehrmann},\
  and\ \citenamefont {Glover}}]{Gehrmann-DeRidder:2005btv}%
  \BibitemOpen
  \bibfield  {author} {\bibinfo {author} {\bibfnamefont {A.}~\bibnamefont
  {Gehrmann-De~Ridder}}, \bibinfo {author} {\bibfnamefont {T.}~\bibnamefont
  {Gehrmann}}, \ and\ \bibinfo {author} {\bibfnamefont {E.~W.~N.}\ \bibnamefont
  {Glover}},\ }\href {\doibase 10.1088/1126-6708/2005/09/056} {\bibfield
  {journal} {\bibinfo  {journal} {JHEP}\ }\textbf {\bibinfo {volume} {09}},\
  \bibinfo {pages} {056} (\bibinfo {year} {2005})},\ \Eprint
  {http://arxiv.org/abs/hep-ph/0505111} {arXiv:hep-ph/0505111} \BibitemShut
  {NoStop}%
\bibitem [{\citenamefont {Daleo}\ \emph {et~al.}(2007)\citenamefont {Daleo},
  \citenamefont {Gehrmann},\ and\ \citenamefont {Maitre}}]{Daleo:2006xa}%
  \BibitemOpen
  \bibfield  {author} {\bibinfo {author} {\bibfnamefont {A.}~\bibnamefont
  {Daleo}}, \bibinfo {author} {\bibfnamefont {T.}~\bibnamefont {Gehrmann}}, \
  and\ \bibinfo {author} {\bibfnamefont {D.}~\bibnamefont {Maitre}},\ }\href
  {\doibase 10.1088/1126-6708/2007/04/016} {\bibfield  {journal} {\bibinfo
  {journal} {JHEP}\ }\textbf {\bibinfo {volume} {04}},\ \bibinfo {pages} {016}
  (\bibinfo {year} {2007})},\ \Eprint {http://arxiv.org/abs/hep-ph/0612257}
  {arXiv:hep-ph/0612257} \BibitemShut {NoStop}%
\bibitem [{\citenamefont {Currie}\ \emph {et~al.}(2013)\citenamefont {Currie},
  \citenamefont {Glover},\ and\ \citenamefont {Wells}}]{Currie:2013vh}%
  \BibitemOpen
  \bibfield  {author} {\bibinfo {author} {\bibfnamefont {J.}~\bibnamefont
  {Currie}}, \bibinfo {author} {\bibfnamefont {E.}~\bibnamefont {Glover}}, \
  and\ \bibinfo {author} {\bibfnamefont {S.}~\bibnamefont {Wells}},\ }\href
  {\doibase 10.1007/JHEP04(2013)066} {\bibfield  {journal} {\bibinfo  {journal}
  {JHEP}\ }\textbf {\bibinfo {volume} {04}},\ \bibinfo {pages} {066} (\bibinfo
  {year} {2013})},\ \Eprint {http://arxiv.org/abs/1301.4693} {arXiv:1301.4693
  [hep-ph]} \BibitemShut {NoStop}%
\bibitem [{\citenamefont {Czakon}(2010)}]{Czakon:2010td}%
  \BibitemOpen
  \bibfield  {author} {\bibinfo {author} {\bibfnamefont {M.}~\bibnamefont
  {Czakon}},\ }\href {\doibase 10.1016/j.physletb.2010.08.036} {\bibfield
  {journal} {\bibinfo  {journal} {Phys. Lett. B}\ }\textbf {\bibinfo {volume}
  {693}},\ \bibinfo {pages} {259} (\bibinfo {year} {2010})},\ \Eprint
  {http://arxiv.org/abs/1005.0274} {arXiv:1005.0274 [hep-ph]} \BibitemShut
  {NoStop}%
\bibitem [{\citenamefont {Gehrmann}\ and\ \citenamefont
  {Stagnitto}(2022)}]{Gehrmann:2022pzd}%
  \BibitemOpen
  \bibfield  {author} {\bibinfo {author} {\bibfnamefont {T.}~\bibnamefont
  {Gehrmann}}\ and\ \bibinfo {author} {\bibfnamefont {G.}~\bibnamefont
  {Stagnitto}},\ }\href {\doibase 10.1007/JHEP10(2022)136} {\bibfield
  {journal} {\bibinfo  {journal} {JHEP}\ }\textbf {\bibinfo {volume} {10}},\
  \bibinfo {pages} {136} (\bibinfo {year} {2022})},\ \Eprint
  {http://arxiv.org/abs/2208.02650} {arXiv:2208.02650 [hep-ph]} \BibitemShut
  {NoStop}%
\bibitem [{\citenamefont {Gehrmann}\ and\ \citenamefont
  {Sch\"urmann}(2022{\natexlab{b}})}]{Gehrmann:2021lwb}%
  \BibitemOpen
  \bibfield  {author} {\bibinfo {author} {\bibfnamefont {T.}~\bibnamefont
  {Gehrmann}}\ and\ \bibinfo {author} {\bibfnamefont {R.}~\bibnamefont
  {Sch\"urmann}},\ }\href {\doibase 10.21468/SciPostPhysProc.7.042} {\bibfield
  {journal} {\bibinfo  {journal} {SciPost Phys. Proc.}\ }\textbf {\bibinfo
  {volume} {7}},\ \bibinfo {pages} {042} (\bibinfo {year}
  {2022}{\natexlab{b}})},\ \Eprint {http://arxiv.org/abs/2110.02617}
  {arXiv:2110.02617 [hep-ph]} \BibitemShut {NoStop}%
\bibitem [{\citenamefont {Chetyrkin}\ and\ \citenamefont
  {Tkachov}(1981)}]{Chetyrkin:1981qh}%
  \BibitemOpen
  \bibfield  {author} {\bibinfo {author} {\bibfnamefont {K.~G.}\ \bibnamefont
  {Chetyrkin}}\ and\ \bibinfo {author} {\bibfnamefont {F.~V.}\ \bibnamefont
  {Tkachov}},\ }\href {\doibase 10.1016/0550-3213(81)90199-1} {\bibfield
  {journal} {\bibinfo  {journal} {Nucl. Phys. B}\ }\textbf {\bibinfo {volume}
  {192}},\ \bibinfo {pages} {159} (\bibinfo {year} {1981})}\BibitemShut
  {NoStop}%
\bibitem [{\citenamefont {Laporta}(2000)}]{Laporta:2000dsw}%
  \BibitemOpen
  \bibfield  {author} {\bibinfo {author} {\bibfnamefont {S.}~\bibnamefont
  {Laporta}},\ }\href {\doibase 10.1142/S0217751X00002159} {\bibfield
  {journal} {\bibinfo  {journal} {Int. J. Mod. Phys. A}\ }\textbf {\bibinfo
  {volume} {15}},\ \bibinfo {pages} {5087} (\bibinfo {year} {2000})},\ \Eprint
  {http://arxiv.org/abs/hep-ph/0102033} {arXiv:hep-ph/0102033} \BibitemShut
  {NoStop}%
\bibitem [{\citenamefont {von Manteuffel}\ and\ \citenamefont
  {Studerus}(2012)}]{vonManteuffel:2012np}%
  \BibitemOpen
  \bibfield  {author} {\bibinfo {author} {\bibfnamefont {A.}~\bibnamefont {von
  Manteuffel}}\ and\ \bibinfo {author} {\bibfnamefont {C.}~\bibnamefont
  {Studerus}},\ }\href@noop {} {\  (\bibinfo {year} {2012})},\ \Eprint
  {http://arxiv.org/abs/1201.4330} {arXiv:1201.4330 [hep-ph]} \BibitemShut
  {NoStop}%
\bibitem [{\citenamefont {Gehrmann}\ and\ \citenamefont
  {Remiddi}(2000)}]{Gehrmann:1999as}%
  \BibitemOpen
  \bibfield  {author} {\bibinfo {author} {\bibfnamefont {T.}~\bibnamefont
  {Gehrmann}}\ and\ \bibinfo {author} {\bibfnamefont {E.}~\bibnamefont
  {Remiddi}},\ }\href {\doibase 10.1016/S0550-3213(00)00223-6} {\bibfield
  {journal} {\bibinfo  {journal} {Nucl. Phys. B}\ }\textbf {\bibinfo {volume}
  {580}},\ \bibinfo {pages} {485} (\bibinfo {year} {2000})},\ \Eprint
  {http://arxiv.org/abs/hep-ph/9912329} {arXiv:hep-ph/9912329} \BibitemShut
  {NoStop}%
\bibitem [{\citenamefont {Duhr}\ and\ \citenamefont
  {Dulat}(2019)}]{Duhr:2019tlz}%
  \BibitemOpen
  \bibfield  {author} {\bibinfo {author} {\bibfnamefont {C.}~\bibnamefont
  {Duhr}}\ and\ \bibinfo {author} {\bibfnamefont {F.}~\bibnamefont {Dulat}},\
  }\href {\doibase 10.1007/JHEP08(2019)135} {\bibfield  {journal} {\bibinfo
  {journal} {JHEP}\ }\textbf {\bibinfo {volume} {08}},\ \bibinfo {pages} {135}
  (\bibinfo {year} {2019})},\ \Eprint {http://arxiv.org/abs/1904.07279}
  {arXiv:1904.07279 [hep-th]} \BibitemShut {NoStop}%
\bibitem [{\citenamefont {Maitre}(2006)}]{Maitre:2005uu}%
  \BibitemOpen
  \bibfield  {author} {\bibinfo {author} {\bibfnamefont {D.}~\bibnamefont
  {Maitre}},\ }\href {\doibase 10.1016/j.cpc.2005.10.008} {\bibfield  {journal}
  {\bibinfo  {journal} {Comput. Phys. Commun.}\ }\textbf {\bibinfo {volume}
  {174}},\ \bibinfo {pages} {222} (\bibinfo {year} {2006})},\ \Eprint
  {http://arxiv.org/abs/hep-ph/0507152} {arXiv:hep-ph/0507152} \BibitemShut
  {NoStop}%
\bibitem [{\citenamefont {Daleo}\ \emph {et~al.}(2010)\citenamefont {Daleo},
  \citenamefont {Gehrmann-De~Ridder}, \citenamefont {Gehrmann},\ and\
  \citenamefont {Luisoni}}]{Daleo:2009yj}%
  \BibitemOpen
  \bibfield  {author} {\bibinfo {author} {\bibfnamefont {A.}~\bibnamefont
  {Daleo}}, \bibinfo {author} {\bibfnamefont {A.}~\bibnamefont
  {Gehrmann-De~Ridder}}, \bibinfo {author} {\bibfnamefont {T.}~\bibnamefont
  {Gehrmann}}, \ and\ \bibinfo {author} {\bibfnamefont {G.}~\bibnamefont
  {Luisoni}},\ }\href {\doibase 10.1007/JHEP01(2010)118} {\bibfield  {journal}
  {\bibinfo  {journal} {JHEP}\ }\textbf {\bibinfo {volume} {01}},\ \bibinfo
  {pages} {118} (\bibinfo {year} {2010})},\ \Eprint
  {http://arxiv.org/abs/0912.0374} {arXiv:0912.0374 [hep-ph]} \BibitemShut
  {NoStop}%
\bibitem [{\citenamefont {Bonino}\ \emph {et~al.}()\citenamefont {Bonino},
  \citenamefont {Gehrmann}, \citenamefont {Marcoli}, \citenamefont
  {Sch\"urmann},\ and\ \citenamefont {Stagnitto}}]{IFfrag}%
  \BibitemOpen
  \bibfield  {author} {\bibinfo {author} {\bibfnamefont {L.}~\bibnamefont
  {Bonino}}, \bibinfo {author} {\bibfnamefont {T.}~\bibnamefont {Gehrmann}},
  \bibinfo {author} {\bibfnamefont {M.}~\bibnamefont {Marcoli}}, \bibinfo
  {author} {\bibfnamefont {R.}~\bibnamefont {Sch\"urmann}}, \ and\ \bibinfo
  {author} {\bibfnamefont {G.}~\bibnamefont {Stagnitto}},\ }\href@noop {} {\
  }\bibinfo {note} {In preparation}\BibitemShut {NoStop}%
\bibitem [{\citenamefont {Gehrmann}\ \emph {et~al.}(2005)\citenamefont
  {Gehrmann}, \citenamefont {Huber},\ and\ \citenamefont
  {Maitre}}]{Gehrmann:2005pd}%
  \BibitemOpen
  \bibfield  {author} {\bibinfo {author} {\bibfnamefont {T.}~\bibnamefont
  {Gehrmann}}, \bibinfo {author} {\bibfnamefont {T.}~\bibnamefont {Huber}}, \
  and\ \bibinfo {author} {\bibfnamefont {D.}~\bibnamefont {Maitre}},\ }\href
  {\doibase 10.1016/j.physletb.2005.07.019} {\bibfield  {journal} {\bibinfo
  {journal} {Phys. Lett. B}\ }\textbf {\bibinfo {volume} {622}},\ \bibinfo
  {pages} {295} (\bibinfo {year} {2005})},\ \Eprint
  {http://arxiv.org/abs/hep-ph/0507061} {arXiv:hep-ph/0507061} \BibitemShut
  {NoStop}%
\bibitem [{\citenamefont {Vermaseren}(2000)}]{Vermaseren:2000nd}%
  \BibitemOpen
  \bibfield  {author} {\bibinfo {author} {\bibfnamefont {J.~A.~M.}\
  \bibnamefont {Vermaseren}},\ }\href@noop {} {\  (\bibinfo {year} {2000})},\
  \Eprint {http://arxiv.org/abs/math-ph/0010025} {arXiv:math-ph/0010025}
  \BibitemShut {NoStop}%
\bibitem [{\citenamefont {H\"oschele}\ \emph {et~al.}(2014)\citenamefont
  {H\"oschele}, \citenamefont {Hoff}, \citenamefont {Pak}, \citenamefont
  {Steinhauser},\ and\ \citenamefont {Ueda}}]{Hoschele:2013pvt}%
  \BibitemOpen
  \bibfield  {author} {\bibinfo {author} {\bibfnamefont {M.}~\bibnamefont
  {H\"oschele}}, \bibinfo {author} {\bibfnamefont {J.}~\bibnamefont {Hoff}},
  \bibinfo {author} {\bibfnamefont {A.}~\bibnamefont {Pak}}, \bibinfo {author}
  {\bibfnamefont {M.}~\bibnamefont {Steinhauser}}, \ and\ \bibinfo {author}
  {\bibfnamefont {T.}~\bibnamefont {Ueda}},\ }\href {\doibase
  10.1016/j.cpc.2013.10.007} {\bibfield  {journal} {\bibinfo  {journal}
  {Comput. Phys. Commun.}\ }\textbf {\bibinfo {volume} {185}},\ \bibinfo
  {pages} {528} (\bibinfo {year} {2014})},\ \Eprint
  {http://arxiv.org/abs/1307.6925} {arXiv:1307.6925 [hep-ph]} \BibitemShut
  {NoStop}%
\bibitem [{\citenamefont {Abele}\ \emph {et~al.}(2021)\citenamefont {Abele},
  \citenamefont {de~Florian},\ and\ \citenamefont {Vogelsang}}]{Abele:2021nyo}%
  \BibitemOpen
  \bibfield  {author} {\bibinfo {author} {\bibfnamefont {M.}~\bibnamefont
  {Abele}}, \bibinfo {author} {\bibfnamefont {D.}~\bibnamefont {de~Florian}}, \
  and\ \bibinfo {author} {\bibfnamefont {W.}~\bibnamefont {Vogelsang}},\ }\href
  {\doibase 10.1103/PhysRevD.104.094046} {\bibfield  {journal} {\bibinfo
  {journal} {Phys. Rev. D}\ }\textbf {\bibinfo {volume} {104}},\ \bibinfo
  {pages} {094046} (\bibinfo {year} {2021})},\ \Eprint
  {http://arxiv.org/abs/2109.00847} {arXiv:2109.00847 [hep-ph]} \BibitemShut
  {NoStop}%
\bibitem [{\citenamefont {Abele}\ \emph {et~al.}(2022)\citenamefont {Abele},
  \citenamefont {de~Florian},\ and\ \citenamefont {Vogelsang}}]{Abele:2022wuy}%
  \BibitemOpen
  \bibfield  {author} {\bibinfo {author} {\bibfnamefont {M.}~\bibnamefont
  {Abele}}, \bibinfo {author} {\bibfnamefont {D.}~\bibnamefont {de~Florian}}, \
  and\ \bibinfo {author} {\bibfnamefont {W.}~\bibnamefont {Vogelsang}},\ }\href
  {\doibase 10.1103/PhysRevD.106.014015} {\bibfield  {journal} {\bibinfo
  {journal} {Phys. Rev. D}\ }\textbf {\bibinfo {volume} {106}},\ \bibinfo
  {pages} {014015} (\bibinfo {year} {2022})},\ \Eprint
  {http://arxiv.org/abs/2203.07928} {arXiv:2203.07928 [hep-ph]} \BibitemShut
  {NoStop}%
\bibitem [{\citenamefont {Borsa}\ \emph {et~al.}(2022)\citenamefont {Borsa},
  \citenamefont {Sassot}, \citenamefont {de~Florian}, \citenamefont
  {Stratmann},\ and\ \citenamefont {Vogelsang}}]{Borsa:2022vvp}%
  \BibitemOpen
  \bibfield  {author} {\bibinfo {author} {\bibfnamefont {I.}~\bibnamefont
  {Borsa}}, \bibinfo {author} {\bibfnamefont {R.}~\bibnamefont {Sassot}},
  \bibinfo {author} {\bibfnamefont {D.}~\bibnamefont {de~Florian}}, \bibinfo
  {author} {\bibfnamefont {M.}~\bibnamefont {Stratmann}}, \ and\ \bibinfo
  {author} {\bibfnamefont {W.}~\bibnamefont {Vogelsang}},\ }\href {\doibase
  10.1103/PhysRevLett.129.012002} {\bibfield  {journal} {\bibinfo  {journal}
  {Phys. Rev. Lett.}\ }\textbf {\bibinfo {volume} {129}},\ \bibinfo {pages}
  {012002} (\bibinfo {year} {2022})},\ \Eprint
  {http://arxiv.org/abs/2202.05060} {arXiv:2202.05060 [hep-ph]} \BibitemShut
  {NoStop}%
\bibitem [{\citenamefont {Abdul~Khalek}\ \emph {et~al.}(2022)\citenamefont
  {Abdul~Khalek}, \citenamefont {Bertone}, \citenamefont {Khoudli},\ and\
  \citenamefont {Nocera}}]{AbdulKhalek:2022laj}%
  \BibitemOpen
  \bibfield  {author} {\bibinfo {author} {\bibfnamefont {R.}~\bibnamefont
  {Abdul~Khalek}}, \bibinfo {author} {\bibfnamefont {V.}~\bibnamefont
  {Bertone}}, \bibinfo {author} {\bibfnamefont {A.}~\bibnamefont {Khoudli}}, \
  and\ \bibinfo {author} {\bibfnamefont {E.~R.}\ \bibnamefont {Nocera}},\
  }\href {\doibase 10.1016/j.physletb.2022.137456} {\bibfield  {journal}
  {\bibinfo  {journal} {Phys. Lett. B}\ }\textbf {\bibinfo {volume} {834}},\
  \bibinfo {pages} {137456} (\bibinfo {year} {2022})},\ \Eprint
  {http://arxiv.org/abs/2204.10331} {arXiv:2204.10331 [hep-ph]} \BibitemShut
  {NoStop}%
\bibitem [{\citenamefont {Goyal}\ \emph {et~al.}(2023)\citenamefont {Goyal},
  \citenamefont {Moch}, \citenamefont {Pathak}, \citenamefont {Rana},\ and\
  \citenamefont {Ravindran}}]{Goyal:2023xfi}%
  \BibitemOpen
  \bibfield  {author} {\bibinfo {author} {\bibfnamefont {S.}~\bibnamefont
  {Goyal}}, \bibinfo {author} {\bibfnamefont {S.-O.}\ \bibnamefont {Moch}},
  \bibinfo {author} {\bibfnamefont {V.}~\bibnamefont {Pathak}}, \bibinfo
  {author} {\bibfnamefont {N.}~\bibnamefont {Rana}}, \ and\ \bibinfo {author}
  {\bibfnamefont {V.}~\bibnamefont {Ravindran}},\ }\href@noop {} {\  (\bibinfo
  {year} {2023})},\ \Eprint {http://arxiv.org/abs/2312.17711} {arXiv:2312.17711
  [hep-ph]} \BibitemShut {NoStop}%
\bibitem [{\citenamefont {Ball}\ \emph {et~al.}(2017)\citenamefont {Ball} \emph
  {et~al.}}]{NNPDF:2017mvq}%
  \BibitemOpen
  \bibfield  {author} {\bibinfo {author} {\bibfnamefont {R.~D.}\ \bibnamefont
  {Ball}} \emph {et~al.} (\bibinfo {collaboration} {NNPDF}),\ }\href {\doibase
  10.1140/epjc/s10052-017-5199-5} {\bibfield  {journal} {\bibinfo  {journal}
  {Eur. Phys. J. C}\ }\textbf {\bibinfo {volume} {77}},\ \bibinfo {pages} {663}
  (\bibinfo {year} {2017})},\ \Eprint {http://arxiv.org/abs/1706.00428}
  {arXiv:1706.00428 [hep-ph]} \BibitemShut {NoStop}%
\bibitem [{\citenamefont {Bertone}\ \emph {et~al.}(2014)\citenamefont
  {Bertone}, \citenamefont {Carrazza},\ and\ \citenamefont
  {Rojo}}]{Bertone:2013vaa}%
  \BibitemOpen
  \bibfield  {author} {\bibinfo {author} {\bibfnamefont {V.}~\bibnamefont
  {Bertone}}, \bibinfo {author} {\bibfnamefont {S.}~\bibnamefont {Carrazza}}, \
  and\ \bibinfo {author} {\bibfnamefont {J.}~\bibnamefont {Rojo}} (\bibinfo
  {collaboration} {APFEL}),\ }\href {\doibase 10.1016/j.cpc.2014.03.007}
  {\bibfield  {journal} {\bibinfo  {journal} {Comput. Phys. Commun.}\ }\textbf
  {\bibinfo {volume} {185}},\ \bibinfo {pages} {1647} (\bibinfo {year}
  {2014})},\ \Eprint {http://arxiv.org/abs/1310.1394} {arXiv:1310.1394
  [hep-ph]} \BibitemShut {NoStop}%
\bibitem [{\citenamefont {Bertone}(2018)}]{Bertone:2017gds}%
  \BibitemOpen
  \bibfield  {author} {\bibinfo {author} {\bibfnamefont {V.}~\bibnamefont
  {Bertone}},\ }\href {\doibase 10.22323/1.297.0201} {\bibfield  {journal}
  {\bibinfo  {journal} {PoS}\ }\textbf {\bibinfo {volume} {DIS2017}},\ \bibinfo
  {pages} {201} (\bibinfo {year} {2018})},\ \Eprint
  {http://arxiv.org/abs/1708.00911} {arXiv:1708.00911 [hep-ph]} \BibitemShut
  {NoStop}%
\end{thebibliography}%

\end{document}